\documentclass[]{interact}

\usepackage{float}
\usepackage{breqn}
\usepackage{epstopdf}
\usepackage[caption=false, skip = 0 pt]{subfig}

\usepackage{color, soul}
\sethlcolor{yellow}

\usepackage[para]{threeparttable}
\usepackage[labelfont=bf]{caption}
\usepackage{adjustbox}
\usepackage{multirow}
\usepackage[numbers,sort&compress]{natbib}
\bibpunct[, ]{[}{]}{,}{n}{,}{,}
\renewcommand\bibfont{\fontsize{10}{12}\selectfont}
\makeatletter
\def\NAT@def@citea{\def\@citea{\NAT@separator}}
\makeatother

\theoremstyle{plain}
\newtheorem{theorem}{Theorem}[section]

\newtheorem{proposition}[theorem]{Proposition}

\theoremstyle{definition}

\theoremstyle{remark}
\newtheorem{remark}{Remark}

\begin{document}
\setlength{\abovedisplayskip}{5pt}
\setlength{\belowdisplayskip}{5pt}

\title{ A Bayesian Mixture Modelling of Stop Signal Reaction Time Distributions }
\author{
\name{Mohsen Soltanifar\textsuperscript{a}\thanks{CONTACT Mohsen Soltanifar. Email: mohsen.soltanifar@mail.utoronto.ca}, Michael Escobar\textsuperscript{b}, Annie Dupuis\textsuperscript{c}, and  Russell Schachar\textsuperscript{d}}
\affil{\textsuperscript{a,b,c} Biostatistics Division, Dalla Lana School of Public Health, University of Toronto, Toronto, M5T 3M7, ON, Canada ;  \textsuperscript{a,c,d} Department of Psychiatry,  The Hospital for Sick Children, 555 University Avenue, Toronto, M5G 1X8, ON, Canada }
}

\maketitle

\begin{abstract}
The distribution of single Stop Signal Reaction Times (SSRT) in the stop signal task (SST) has been modeled with two general methods: a nonparametric method by Hans Colonius (1990); and a Bayesian parametric method by Dora Matzke, Gordon Logan and colleagues (2013). These methods assume an equal impact of the preceding trial type (go/stop) in the SST trials on the SSRT distributional estimation without addressing the relaxed assumption. This study presents the required model by considering a two-state mixture model for the SSRT distribution. It then compares the Bayesian parametric single SSRT and mixture SSRT distributions in the usual stochastic order at the individual and the population level under ex-Gaussian(ExG) distributional format. It shows that compared to a single SSRT distribution, the mixture SSRT distribution is more diverse, more positively skewed, more leptokurtic, and larger in stochastic order. The size of the results' disparities also depends on the choice of weights in the mixture SSRT distribution. This study confirms that mixture SSRT indices as a constant or distribution are significantly larger than their single SSRT counterparts in the related order. This result offers a vital improvement in the SSRT estimations.
\end{abstract}


\section{Introduction}

Inhibition refers to the ability to suppress actively, interrupt or delay an action \cite{I1}. Inhibition itself is a crucial dimension of executive control, which on its own is required for an organism to adjust behavior according to changing conditions; this could be assessing inappropriateness of the current course of thought and action, changing goals, and changing world, \cite{I2,I3,I4}. Response inhibition is the ability to stop responses that are no longer appropriate \cite{I3}. Examples of response inhibition in daily life include braking quickly when driving into an intersection while another vehicle is running through a red light \cite{I4}. Two paradigms have been suggested to study response inhibition empirically in a laboratory setting: The Go/No-go task and the stop-signal task (SST). The later is widely used, \cite{I1,I5}. The stop-signal paradigm includes two response tasks: the go task and the stop task(e.g., stop 25\% of the time). In go trials, the go reaction time (GORT) is the response to the stimulus such as “X" and “O" presented on the computer screen.  In stop trials,  the stop-signal reaction time (SSRT) is the unobserved latency of the stopping response in the brain upon observing the stop signal (e.g., an auditory tone such as “beep").  The stop signal is presented to the participant after the passage of some time called the stop signal delay  \cite{I6,I7}. Often, the adjustment of stop-signal delays (SSD or $T_d$)  is made by the more reliable tracking method in which, depending on the previous trial's success or failure, the $T_d$ is increased or decreased by 50 ms to achieve 50\% overall successful inhibition at the end of the paradigm.  In the go trials and stop signal response trials, the observed reaction times and the unobserved latency of the stopping response(i.e., GORT, SRRT, and SSRT, respectively) are measured milliseconds. In young adults trying to stop continuous actions, such as typing, the SSRT is close to 200 ms \cite{I8}. \par
\begin{figure}[H]
\begin{center}
\includegraphics[totalheight=10.0 cm, width=14 cm]{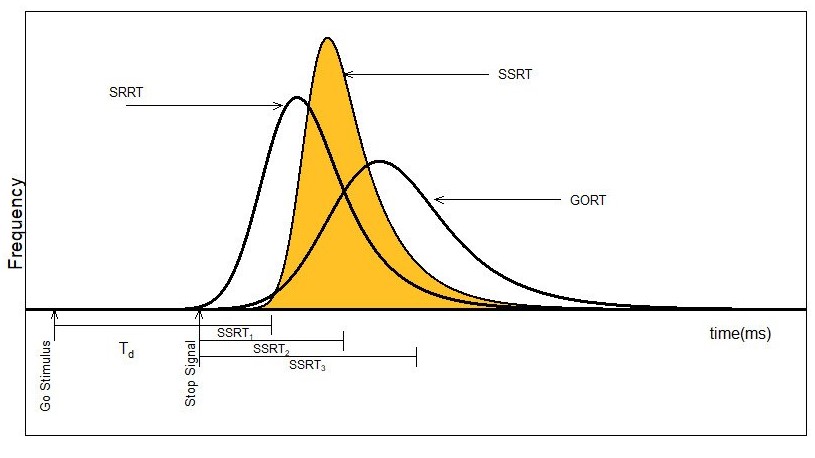}
\caption{Graphical representation of the complete horse race model[\cite{I3}, modification] : GORT: go reaction times, SRRT: Signal respond reaction times,  SSRT: Stop Signal reaction times; $T_d$: stop signal delay(SSD)}
\end{center}
\end{figure}
Several models have been proposed to evaluate and describe response inhibition in the stop-signal paradigm, including the deadline model, the independent horse race model, the interactive horse race model, and the Hanes-Carpenter model \cite{I9,I10,I11,I12}. In this study, the independent horse race model is considered. It provides a theoretical framework in which researchers can measure the Stop Signal Reaction Times (SSRT) and its associated factors \cite{I13}. There are two types of the horse race model: the independent model with constant SSRT index and the complete independent model with non-constant SSRT distribution; in this study, we focus on the second model(See Figure 1).  SSRT measurements have been a critical tool used by psychopathologists to make inferences about a patient's ability to inhibit thought and action (i) on the spectrum of clinical groups (e.g., ADHD, OCD, autism, schizophrenia ); and, (ii) across various tasks and experimental conditions \cite{I13}. SSRT measurement's precise estimation affects such inferences profoundly. \par 
There are several estimation methods of SSRT depending on two contexts in the SST literature: (i)  as a constant index, or (ii) as a non-constant random variable. Within each context, these methods mentioned below refined the earlier proposed methods given their associated contexts.\par  
There are four estimation methods of SSRT as a constant index, including the mean crude method, the Logan 1994 integration method\cite{I3}, the weighted method\cite{I14} and the time series based state-space method\cite{I15}. Given a subject with go reaction time GORT random variable in the go trials with quantile function $Q_{GORT}$, $n$ stop signal delays  $T_d$ ,  and the probability of successful inhibition(SI). Then, the first couple of the point indices of constant SSRT in the entire SST cluster are defined as:\par 
\begin{eqnarray}
SSRT_{Crude}^{c}&=&\overline{GORT}-\overline{T_d}\\
SSRT_{Logan1994}^{c}&=&Q_{GORT}(1-P(SI|\overline{T_d}))-\overline{T_d}.
\end{eqnarray}
Given higher reliability and less bias in the second index versus the first index, the second index has been recommended as the plausible index of constant SSRT \cite{I16}. The third and fourth indices are essentially improvements of the second index under their associated contexts. For the third point index of constant SSRT, partitioning the entire SST cluster into two clusters of type-A SST cluster (trials following a go trial) and type-B SST cluster (trials following a stop trial) and calculating trial-type related Logan 1994 SSRT as $SSRT_A$ and $SSRT_B$ with corresponding weights $W_{A}= \#\text{Type A Stop}/\#\text{Total Stop},$  $W_B=1-W_A,$ the third point index of constant SSRT is defined as:\par 
\begin{equation}
SSRT_{Weighted}^{c}=W_{A}.SSRT_{A}+W_{B}.SSRT_{B}.
\end{equation}
Finally, one may transform raw tri-variate SST time series data to tri-variate state-space time series data using the missing data EM algorithm.  Given the EM algorithm  log-normally distributed outputs $GORT^{ss.ln}$ and $T_{d}^{ss.ln},$ the fourth point index of constant SSRT is defined as: \par 
\begin{equation}
SSRT_{SS.Logan1994}^{c} = Q_{GORT^{ss.ln}}(1-P(SI|\overline{T_{d}^{ss.ln}}))-\overline{T_{d}^{ss.ln}}.
\end{equation}
Several researchers have shown that merely focusing on measures of central tendency in reaction times RT, including SSRT, gives insufficient information regarding the data's nature. For instance, different clinical groups may have the same mean reaction times. However, the shape of their distributions differs in several aspects. The differences are in their tails as seen in an ADHD group compared to the controls \cite{I17} or their domain of variance in a schizophrenia group versus control \cite{I18}. These observations lead the researchers to study the entire SSRT distribution (Figure 1).\par 
There are two main methods to estimate SSRT as a random variable, including Colonius’s nonparametric method \cite{I19} and the Bayesian parametric method \cite{I20, I21}.  The first method retrieves the cumulative distribution function of SSRT given several components as follows: (i) go reaction times GORT in go trials with density $f_{GORT}$, (ii) signal respond reaction times SRRT in the failed stop trials with density $f_{SRRT}$, (iii) n stop signal delays  $T_d,$ and (iv) probability of successful inhibition(SI). The cumulative distribution function is calculated as:    \par  
\begin{equation}
F_{SSRT}(t)=1-(1-P(SI|T_{d})).(\frac{f_{SRRT}(t+T_d|T_d)}{f_{GORT}(t+T_d)}),\ \ 0<t,T_d<\infty.
\end{equation}
Although the first method theoretically gives the entire non-parametric distribution of SSRT, it cannot be implemented for empirical data in practice.  It requires an unrealistically high number of trials for accurate estimations. In the case of such estimations for simulated data, it has underestimated the mean of SSRT and overestimated the variance of SSRT \cite{I4, I22}.  These observations lead researchers to propose the second method of estimation of SSRT in the  Bayesian context under given parametric distributional assumptions for the involved GORT, SRRT, and SSRT in the SST data, \cite{I20}. The Bayesian Parametric Approach (BPA) presents a novel parametric approach to estimate the entire distribution of SSRT, which applies to real data with a low trial number \cite{I21}. Depending on the individual or hierarchical data, the BPA estimates parameters of the SSRT distribution distinctively. The estimation is done separately for each participant (called individual BPA or IBPA) or is done successively for each participant and, then the entire population(called hierarchical BPA or HBPA) \cite{I23,I24}. The BPA is based on the idea of censoring. Here, the censored observation on the right side of the censoring point $(T_{d}+SSRT)$ in the signal inhibit stop trials are omitted, but their number is known. It also assumes the fundamental assumption of the independent horse race model: the GORT and SSRT are independent distributions \cite{I21}. Assuming that the distribution of GORT and SSRT follows a parametric form, such as Ex-Gaussian, Ex-Wald, Shifted Wald, Shifted Lognormal or Shifted Weibull with parameter $\theta=(\theta_1,\theta_2,\theta_3),$ the goal of the BPA is to estimate simultaneously the parameters $\theta_{go},\theta_{stop}.$ First, given $G$  go trials with go RT times $t=(t_g )_{g=1}^{G}$, the log likelihood is given by:\par 
\begin{equation}
Log(L(\theta_{go}|t))=\sum_{g=1}^{G} log(f_{GO}(t_g|\theta_{go}))
\end{equation}
Second, given S stop trials, including R signal respond stop trials with go SRT times $t=(t_r)_{r=1}^{R}$,  and $I$ signal inhibit trials with delays $T_d=(t_{d_s})_{s=1}^{S} , (S=R+I),$  the log-likelihood of is given by:\par 
\begin{eqnarray}
Log(L(\theta_{go},\theta_{stop}|t,T_d))&=& \sum_{r=1}^{R} log(f_{go}(t_r|\theta_{go}))+
\sum_{r=1}^{R}log(1-F_{stop}(t_r-T_{d_r}|\theta_{stop}))\nonumber\\ 
&+& \sum_{i=1}^{I} log (\int_{0}^{\infty} (1-F_{go}(t|\theta_{go}))\times f_{stop}(t-T_{d_{S-I+i}})dt).
\end{eqnarray}
Finally, although standard MLE methods can be applied to estimates, posterior estimates  $\theta_{go},\theta_{stop},$  given BPA intends to handle both individual and hierarchical cases.  It applies Markov Chain Monte Carlo (MCMC) sampling to estimate them \cite{I20, I21}. The software used for the computation includes WINBUGS \cite{I20} and Bayesian Ex-Gaussian Estimation of Stop-Signal RT distributions (BEESTS) \cite{I21}. Several studies have used the BPA approach in estimating SSRT distribution parameters, for the case of Ex-Gaussian distribution assumption with $\theta=(\mu,\sigma,\tau).$ For example, it has been shown that more practice in stop trials corresponds to lower estimated $\mu$ and higher estimated $\tau$, for the SSRT distribution, \cite{I25}.  Next, the  BPA approach has shown that tyrosine consumption corresponds with lower estimated $\mu$ for the SSRT distribution \cite{I26}. Finally, the mixture BPA approach has been used to show the existence of trigger failures\footnote{Trigger Failure(TF) refers to the situation in which the participant fails to correctly dignose and interpret the stop signal leading to his inability to attempt to inhibit the ongoing go process, \cite{I26-2}.} on stop-signal performance in healthy control participants in two studies of inhibition deficiency in schizophrenia \cite{I27}.\par 
However, as mentioned by Logan \cite{I4} there is little known about the inhibition's aftereffects and the type of questions of interest. One related unanswered question is whether there exist any aftereffects of the non-inhibited(e.g., go) trials and inhibited (e.g., stop)trials on inhibition and in case of affirmative answer how to measure SSRT as a random variable. Here, in both the nonparametric and Bayesian parametric methods mentioned above, there is an implicit assumption. The assumption is about the aftereffects of go trials and stop trials in SST data; that is, the impact of the preceding trial type, either go or stop, on the current stop trial SSRT is assumed to be the same. Most of the SST literature has taken this assumption for granted. To the best of the authors' knowledge, few studies have dealt with this question and estimated the SSRT distribution when this assumption is relaxed. Some studies have shown that after a go trial, the participants have a lower go reaction time GORT versus after the stop trial \cite{I28}. This phenomenon implies that the GORT  distribution after each type of trial(go/stop) will differ, impacting the participant's ability to stop after each trial type \cite{I29}. There are only two studies in the SST literature that partially answered this question when SSRT is considered as constant index \cite{I14, I15}. Here, it was shown when considering SST data in a longitudinal context   $SSRT_{Weighted}^{c}>SSRT_{Logan1994}^{c};$ and when considering SST data in a missing time series context  $SSRT_{(SS.Logan1994)}^{c}>SSRT_{Logan1994}^{c}.$ Both studies' results were valid for the empirical SST data and the simulated SST data. Given that constant SSRT index can be considered as a degenerate random variable, these results partially shed light on the proposed question for the case of a non-degenerate SSRT random variable. However, it is still unknown in which order context these comparisons over random variables can be conducted; additionally, in which mechanism the pairwise comparisons of the involved paired sets of random variables is conducted.\par 
This study offers an estimation of the SSRT distribution given the violated assumption of equal impact of the preceding trial type(go/stop) on the current stop trial SSRT distribution. It uses the notion of two-state mixtures \cite{I30} and proposes parametric mixture Bayesian modeling on the entire SST data set. The study's outline is as follows: First, as in \cite{I14} for each participant, the overall empirical SST data is partitioned into type A cluster SST data and type B cluster SST data. Using the IBPA method, the fitted SSRT Ex-Gaussian parameters are calculated for the cluster type SSRT distributions and the single SSRT distribution. The study's empirical data provides an example of the violated assumption. Second, a mixture SSRT random variable is introduced as a natural generalization of two cases: (i) its constant index $SSRT_{Weighted}^{c}$ counterpart in equation (3); and, (ii) its Bayesian parametric form under the ExGaussian distributional assumption.  Then, considering the mean of posterior parameters as their point estimations, the key descriptive and shape statistics of the mixture SSRT $(SSRT_{Mixture})$ are compared with those of type A SSRT $(SSRT_{A})$, type B SSRT $(SSRT_{B}),$ and the single SSRT $(SSRT_{Single})$. Third, we compare the involved pairs of distributions in usual stochastic order $(<_{st})$ at the individual and population level. The population-level comparisons use our proposed Two-Stage Bayesian Parametric Approach (TSBPA) and our proposed Paired Samples Parametric Distribution Test (PSPDT).  Finally, the earlier comparisons are repeated and discussed in terms of the involved weights in the definition of proposed mixture SSRT $(SSRT_{Mixture}).$\par 

\section{Materials \& Methods}

\subsection{The Data and Study Design}

This study's data and design are previously described at the \cite{M1}. That study included 16,099 participants aged 6 to 19 years old and was conducted at the Ontario Science Center in Toronto, Canada, between June 2009 and September 2010. Each participant completed four blocks of 24 trials with a random 25\% stop signal trials in each block. There were 96 trials in total(24 stop signals; and 72 go trials). Every go trial began with a   500 ms fixation point followed by a stimulus: An O or X presented for 1000 ms in the center of a computer screen. With an initial stop signal delay $T_{d}$  of 250 ms after the go stimulus, each stop trial included an audio stop signal cue presented through headphones to the participant in the context of the tracking method. \par 

\subsection{The Sample and Variables}

\subsubsection{Cluster Type SST data}

A random sample of  44 participants was selected for further analysis. The entire stop signal task data for each participant was partitioned to cluster types, as shown in Table 1, \cite{I14}: (i) type-A stop signal task data (all trials preceded by go trials) and (ii) type-B stop signal task data (all trials preceded by stop trials). These participants each had a minimum of 10 type B stop trials.

\begin{table}[H]
   \caption{Partition of stop task (SST) data to Type A SST data and Type B SST data given previous trial type (go/stop), \cite{I14}}  
   \label{tab:example_multirow}
   \small 
   \centering 
   \begin{tabular}{lcccr} 
   Data &  \multicolumn{2}{c}{ } & \multicolumn{2}{c}{ Previous Trial}\\ 
   \midrule
     &  & & Go & Stop \\ 
   \midrule
   \multirow{2}{*}{Current Trial} & &    Go & $Go_A$ & $Go_B$  \\
                                  &  & Stop & $Stop_A$ & $Stop_B$ \\
    \midrule
   \end{tabular}
\end{table}

For each participant, four types of SST data clusters were considered: Type -A cluster, Type-B cluster, Type-S Single cluster (entire single SST data), and type-M Mixture cluster (the composition of type-A and type-B clusters). Using IBPA, the corresponding  Ex-Gaussian SSRTs’ parameters $\theta=(\mu,\sigma,\tau)$ were calculated as described in the next subsections.\par  

\subsubsection{Ex-Gaussian Random Variable}

Heathcote (1996), \cite{M2} formulated the Ex-Gaussian (ExG) distribution with parameters $(\mu,\sigma,\tau)$ with density given by:\par 
\begin{equation}
f_{ExG}(t|\mu,\sigma,\tau)= \frac{1}{\tau} exp(\frac{\mu-t}{\tau}+\frac{\sigma^2}{2\tau^2})\times \Phi(\frac{\mu-t}{\sigma}-\frac{\sigma}{\tau}):\\ \sigma,\tau>0, -\infty<t<\infty
\end{equation}

where $\Phi$ is the standard normal cumulative distribution function. The first four non-central moments are given by:\par 

\begin{eqnarray}
E(ExG)&=& \mu+\tau, \nonumber\\
E(ExG^2)&=& \mu^2+2\mu\tau+\sigma^2+2\tau^2,\nonumber\\
E(ExG^3)&=& \mu^3+3\mu\sigma^2+6\mu\tau^2+3\mu^2\tau+3\sigma^2\tau+6\tau^3, \nonumber\\
E(ExG^4)&=& \mu^4+4\mu^3\tau+6\mu^2\sigma^2+12\mu^2\tau^2+24\mu\tau^3+12\mu\sigma^2\tau+3\sigma^4+12\sigma^2\tau^2+24\tau^4.\nonumber\\
&&
\end{eqnarray}
Finally, this random variable is right-skewed and leptokurtic with the following variance, skewness, and kurtosis shape statistics:\par  
\begin{eqnarray}
Var(ExG)&=&\sigma^2+\tau^2\nonumber\\
\gamma_{ExG}&=& 2(1+\sigma^2\tau^{-2})^{\frac{-3}{2}}\nonumber\\
\kappa_{ExG}&=& 3\bigg(\frac{1+2\sigma^{-2}\tau^{2}+3\sigma^{-4}\tau^{4}}{(1+\sigma^{-2}\tau^2)^2}\bigg).
\end{eqnarray}

\subsubsection{Mixture SSRT Random Variable}

 Given single SSRT by $SSRT_{S}$, type-A SSRT by  $SSRT_{A}$  type-B SSRT by $SSRT_{B}$,  and  $W_A \sim Bernoulli (W_{A}^{c})$  where the type A trial type weight $W_{A}^{c}$  is given by  $W_{A}^{c}=\#\text{Type A Stop}/\#\text{Total Stop},$ $W_{B}^{c}=1-W_{A}^{c},$ the  Single SSRT and Mixture SSRT random variables were defined as follows\footnote{Note that with notation SSRT for random variable SSRT and $SSRT^c$ for constant SSRT estimated with frequentist methods, we have: $E(SSRT)=SSRT^c$. Consequently, definitions in equation (11) are natural generalizations of constant SSRT estimations with frequentist methods \cite{I14} to general non-constant random variables.}\footnote{Here onward, $W_A$ given the context is either a Bernoulli random variable or a constant number $W_{A}^{c}$ defined as above.}:\par 
\begin{eqnarray}
SSRT_{Single}&=& SSRT_{S},\nonumber\\
SSRT_{Mixture}&=& W_A\times SSRT_{A}+W_{B}\times SSSRT_{B}.
\end{eqnarray}
In the Bayesian context and using IBPA  and under Ex-Gaussian parametric assumption, we have         
$SSRT_S\sim ExG(\theta_S=(\mu_S,\sigma_S,\tau_S)),$ $ SSRT_A\sim ExG(\theta_A=(\mu_A,\sigma_A,\tau_A)) $ and $SSRT_B\sim ExG(\theta_B=(\mu_B,\sigma_B,\tau_B)),$ where the parameter point estimations  are the means of the associated posterior distributions in IBPA. The Bayesian Mixture Ex-Gaussian SSRT model can be formulated as follows. The priors in the IBPA have uninformative uniform distribution and their own chosen parameters $(\alpha,\beta)$ are based on the positive ranges of parameters $(\mu,\sigma,\tau)$ of the associated ExG distribution. Figure 2 presents the model using plate notation. Here, we have:\newline\\
\noindent$K=2:$ Number of cluster types,\newline\\
$N=96:$ Number of trials in SST data,\newline\\
$\theta_{i}=(\mu_i,\sigma_i,\tau_i):$ Parameters of Ex-Gaussian SSRT distribution of the ith cluster $(i=1:A; i=2:B),$ \newline\\
$\mu_i\sim U[\alpha_1,\beta_1], (i=1:A; i=2:B):$ Here: $\alpha_1=10, \beta_1=2,000,$\newline\\
$\sigma_i\sim U[\alpha_2,\beta_2], (i=1:A; i=2:B):$ Here: $\alpha_2=10, \beta_2=2,000,$\newline\\ 
$\tau_i\sim U[\alpha_3,\beta_3], (i=1:A; i=2:B):$ Here: $\alpha_3=10, \beta_3=2,000,$\newline\\
$\phi=(\phi_1,\phi_2):$ Prior Probability of clusters $(\phi_1=W_{A},\phi_2=W_{B}),$\newline\\
$z_{i}\sim Bernoulli(W):\ \ W=W_{A},$ \newline\\
$x_{i}:$ ith SST trial,\newline\\
$x_{i}|_{stop}\sim ExG(\mu_{z_i},\sigma_{z_i},\tau_{z_i}).$\newline\\
\begin{figure}[H]
\begin{center}
\includegraphics[scale = 0.1]{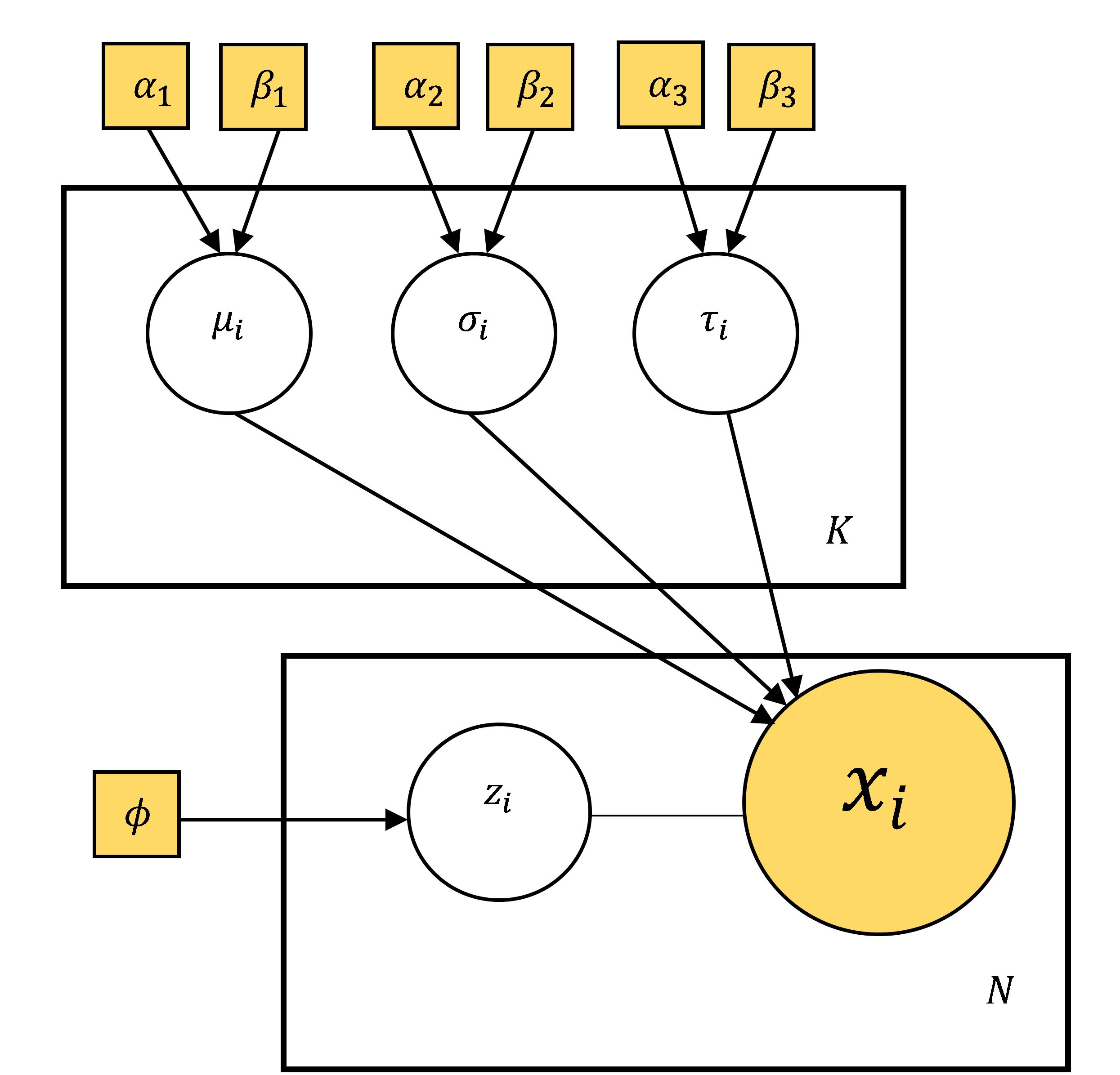}
\vspace*{\floatsep}
\caption{Bayesian Mixture Ex-Gaussian SSRT model using plate notation. Filled in shapes indicate known values.}
\end{center}
\end{figure} 
The first four moments of the Mixture SSRT are as follows:
\begin{equation}
E(SSRT_{Mixture}^{k})=W_{A} E(SSRT_{A}^{k})+W_{B} E(SSRT_{B}^{k}):\ \ 1\leq k\leq 4.
\end{equation}
Consequently, the variance, the skewness and the kurtosis of the Mixture SSRT are computed by:\newpage 
\begin{align}
& Var(SSRT_{Mixture})&=&W_{A}E(SSRT_{A}^2)+(1-W_{A})E(SSRT_{B}^2)\nonumber\\
& &&-(W_{A}E(SSRT_{A})+(1-W_A)E(SSRT_{B}))^2, \nonumber\\
& \gamma_{SSRT_{Mixture}}&=& \frac{1}{Var^{\frac{3}{2}}(SSRT_{Mixture})} \bigg(E(SSRT_{Mixture}^3)\nonumber\\
& &&-3E(SSRT_{Mixture})E(SSRT_{Mixture}^2)+2E^{3}(SSRT_{Mixture}) \bigg),\nonumber\\
& \kappa_{SSRT_{Mixture}}&=& \frac{1}{Var^{2}(SSRT_{Mixture})} \bigg(E(SSRT_{Mixture}^4)-4E(SSRT_{Mixture})E(SSRT_{Mixture}^3) \nonumber\\
& &&+6E^2(SSRT_{Mixture})E(SSRT_{Mixture}^2)-3E^4(SSRT_{Mixture})\bigg).\nonumber\\
& &&
\end{align}
\begin{remark}
Using new equation (11) for SSRT,  Colonious’s proposed nonparametric method for retrieving the entire SSRT CDF for given type-A weight $W_A,$ type A delay $T_{dA}$, type-A signal respond density $f_{SRRTA}$, type-A GORT density $f_{GORTA}$, type-A probability of successful inhibition $P(SI|T_{dA})$;  and, the corresponding type B information yields the following mixture form:\par 
\end{remark} 
\begin{eqnarray}
F_{SSRT}(t)&=&1-W_{A}\bigg((1-P(SI|T_{dA})).(\frac{f_{SRRT}(t+T_{dA}|T_{dA})}{f_{GORT}(t+T_{dA})})\bigg)\nonumber\\
&&-W_{B}\bigg((1-P(SI|T_{dB})).(\frac{f_{SRRT}(t+T_{dB}|T_{dB})}{f_{GORT}(t+T_{dB})})\bigg),\nonumber\\
&&\hspace{5 cm} 0<t,T_{dA},T_{dB}<\infty.
\end{eqnarray}
\begin{remark}
The mixture modelling for SSRT proposed here can be applied with other non-Ex-Gaussian  parametric RT distributions such as Ex-Wald, Wald, \cite{M3} Gamma, Weibull, and Lognormal \cite{M4,M5} with the required modifications in estimations.\par   
\end{remark}
 
\subsection{Statistical Analysis}

For each participant IBPA under Ex-Gaussian parametric distribution was run three times: one for its associated cluster type-A, cluster type-B and single type-S SST data (a total of 132 times). We then calculated the mean posterior estimates of $\theta_S=(\mu_S,\sigma_S,\tau_S),$ $\theta_A=(\mu_A,\sigma_A,\tau_A),$  and $\theta_B=(\mu_B,\sigma_B,\tau_B).$ Then, the parameters, the descriptive statistics and the shape statistics for type-A SSRT $(SSRT_A),$ type-B SSRT $(SSRT_B),$ type-S single SSRT $(SSRT_{Single})$ and type-M Mixture SSRT $(SSRT_{Mixture})$ were calculated. The next steps of the analysis depended to the context and procedure described in the following. 

\subsubsection{Comparisons Context: Real Numbers and Random Variables}

Two sets of comparisons were conducted: (i) within a real numbers contexts; and (ii) within a real-valued random variables context.  For the first set of comparisons,  paired t-test (PROC TTEST, ‘SAS/STAT’ software version 9.4, \cite{M6}) were conducted. These comparisons were made for the Ex-Gaussian distribution's fitted parameters, the descriptive summary statistics, and the shape statistics in the usual real numbers order $(<)$ across cluster types. For the second set of comparisons,  the two samples Kolmogorov Smirnov(KS) tests ( ks.test package stats, ‘R’ software version R.3.4.3, \cite{M7} ) were conducted. These tests were conducted under the assumption of 96 points for the involved random variables CDFs to compare the SSRT random variables in usual stochastic order $(<_{st})$ across cluster types. Such test at the individual is:
\begin{eqnarray}
H_{0}&:&SSRT_{Single}(\overrightarrow{\theta_S})=_{st}SSRT_{Mixture}(\overrightarrow{\theta_M}),\nonumber\\
H_{1}&:&SSRT_{Single}(\overrightarrow{\theta_S})\neq_{st}SSRT_{Mixture}.(\overrightarrow{\theta_M}) 
\end{eqnarray}

\subsubsection{Comparisons Procedure: Random Variables}

Given two sets of stop signal reaction times distributions $\lbrace SSRT_{Single}(\theta_{S_k} )\rbrace_{k=1}^{44}$ and $\lbrace SSRT_{Mixture}(\theta_{M_k})\rbrace_{k=1}^{44}$, our problem of interest was  an overall comparison between these two groups of distributions in usual stochastic order $<_{st}$ \cite{M8}. Our proposed problem was dealt with in two steps as follows: \newline\\ 
\textbf{Step(1): Two-Stage Bayesian Parametric Approach(TSBPA\footnote{This proposed analysis is neither completely hierarchical Bayesian analysis nor completely conventional meta-analysis. It has components of both methods. On the one hand, it has two separates one-stage Bayesian analyses. On the other hand, it calculates overall population-level estimates in the second analysis with consideration of non-zero correlations.})} 

Referring  to equation (11), we define overall SSRT distributions per single S cluster type and mixture M cluster type as the following:\par 
\begin{eqnarray}
SSRT_{O.Single}(\overrightarrow{\theta_S})&=&SSRT_{Single}(\overline{\theta_S}),\nonumber\\
SSRT_{O.Mixture}(\overrightarrow{\theta_M})&=&SSRT_{Mixture}(\overline{\theta_M}).
\end{eqnarray}
where  $\overrightarrow{\theta_S}=\overline{\theta_S}=\overline{\theta_T}=(\overline{\mu_T},\overline{\sigma_T},\overline{\tau_T})$ and $\overrightarrow{\theta_M}=\overline{\theta_M}=(\overline{W_{A}},\overline{\theta_A},\overline{\theta_B})$ with $\overline{W_{A}}=\sum_{k=1}^{44}W_{A_k}/44$ and $\overline{\theta_A}=(\overline{\mu_A},\overline{\sigma_A},\overline{\tau_A}), \overline{\theta_B}=(\overline{\mu_B},\overline{\sigma_B},\overline{\tau_B})$   being computed  by a Two Stage Bayesian Parametric Approach(TSBPA) method described as follows.\par 
In the TSBPA (See Figure 3), the data, the priors and the posterior estimations are considered as below \cite{M9,M10,M11}. We conduct the first stage with 3 chains, 5,000 burn in out of 20,000 simulations in BEESTS 2.0 software. Then we consider the mean of posterior estimates $\mu_{stop},\sigma_{stop},\tau_{stop}$ as their point estimates  $E(\mu_{stop}|x)\rightarrow \mu_{stop}, E(\sigma_{stop}|x)\rightarrow \sigma_{stop},  E(\tau_{stop}|x)\rightarrow \tau_{stop}$ in the second stage of meta-analysis. We conduct this stage with 3 chains with 5,000 burn in out of 100,000 simulations in WINBUGS1.4 software \cite{M12}. Finally, we consider the mean of posterior estimates  $\mu_{\mu_{stop}}, \mu_{\sigma_{stop}}, \mu_{\tau_{stop}}$ in the second stage as estimates of $\overrightarrow{\theta_S}=\overline{\theta_S}=\overline{\theta_T}=(\overline{\mu_T},\overline{\sigma_T},\overline{\tau_T})$ respectively for the case of overall data S. We repeat this process for the case of type A SST data and type B SST data for estimation of $\overline{\theta_A}=(\overline{\mu_A},\overline{\sigma_A},\overline{\tau_A}), \overline{\theta_B}=(\overline{\mu_B},\overline{\sigma_B},\overline{\tau_B})$, respectively. 

\begin{table}[H]
\centering 
\label{table-nonumber}
\begin{threeparttable}[b]
\begin{tabular}{ll} 
Stage(1)&\\
&\\ 
Data & Individual Priors\\
 & \\  
$GORT \sim ExG(\mu_{go},\sigma_{go},\tau_{go})$&  \\
$SRRT\sim ExG(\mu_{go},\sigma_{go},\tau_{go},\mu_{stop},\sigma_{stop},\tau_{stop},SSD)I_{[1,1000]}^{+}$& $\mu_{go},\sigma_{go},\tau_{go}\sim U[10,2000]$\\  
$SSRT\sim ExG(\mu_{go},\sigma_{go},\tau_{go},\mu_{stop},\sigma_{stop},\tau_{stop},SSD)I_{[1,1000]}^{+}$& $\mu_{stop},\sigma_{stop},\tau_{stop}\sim U[10,2000]$ \\
\end{tabular}
\end{threeparttable}
\end{table}  

\begin{table}[!htbp]
\centering 
\label{table-nonumber}
\begin{threeparttable}[b]
\begin{tabular}{ll} 
Stage(2)&\\
&\\ 
Data &  Priors\\
 & \\  
$(\mu_{stop},\sigma_{stop},\tau_{stop})'\sim MVN(M_{3\times1},\sum_{3\times3})$ &
 $\rho_{\mu,\sigma},\rho_{\mu,\tau},\rho_{\sigma,\tau}\sim U[-0.99,+0.99]$ \\
$\mu_{stop}\sim N(\mu_{\mu_{stop}},\sigma_{\mu_{stop}}^2)$ &
$\mu_{\mu_{stop}},\beta_{20},\beta_{30}\sim N(0,1000)$ \\
$\sigma_{stop}|\mu_{stop}\sim N(\mu_{\sigma_{stop}},\sigma_{\sigma_{stop}}^2)$ &
$\sigma_{\mu_{stop}},\sigma_{\sigma_{stop}},\sigma_{\tau_{stop}}\sim N(0,10)I_{[0,1000]}^{+}$ \\
$\tau_{stop}|(\mu_{stop},\sigma_{stop})\sim N(\mu_{\tau_{stop}},\sigma_{\tau_{stop}}^2)$ &\\
$\mu_{\sigma_{stop}i}=\beta_{20}+\beta_{21}.\mu_{stop}i$ &\\
$\mu_{\tau_{stop}i}=\beta_{30}+\beta_{31}.\mu_{stop}i+\beta_{32}.\sigma_{stop}i$ & \\
\end{tabular}
\end{threeparttable}
\end{table}

\begin{figure}[H]
\begin{center}
\includegraphics[totalheight=15 cm, width=12 cm]{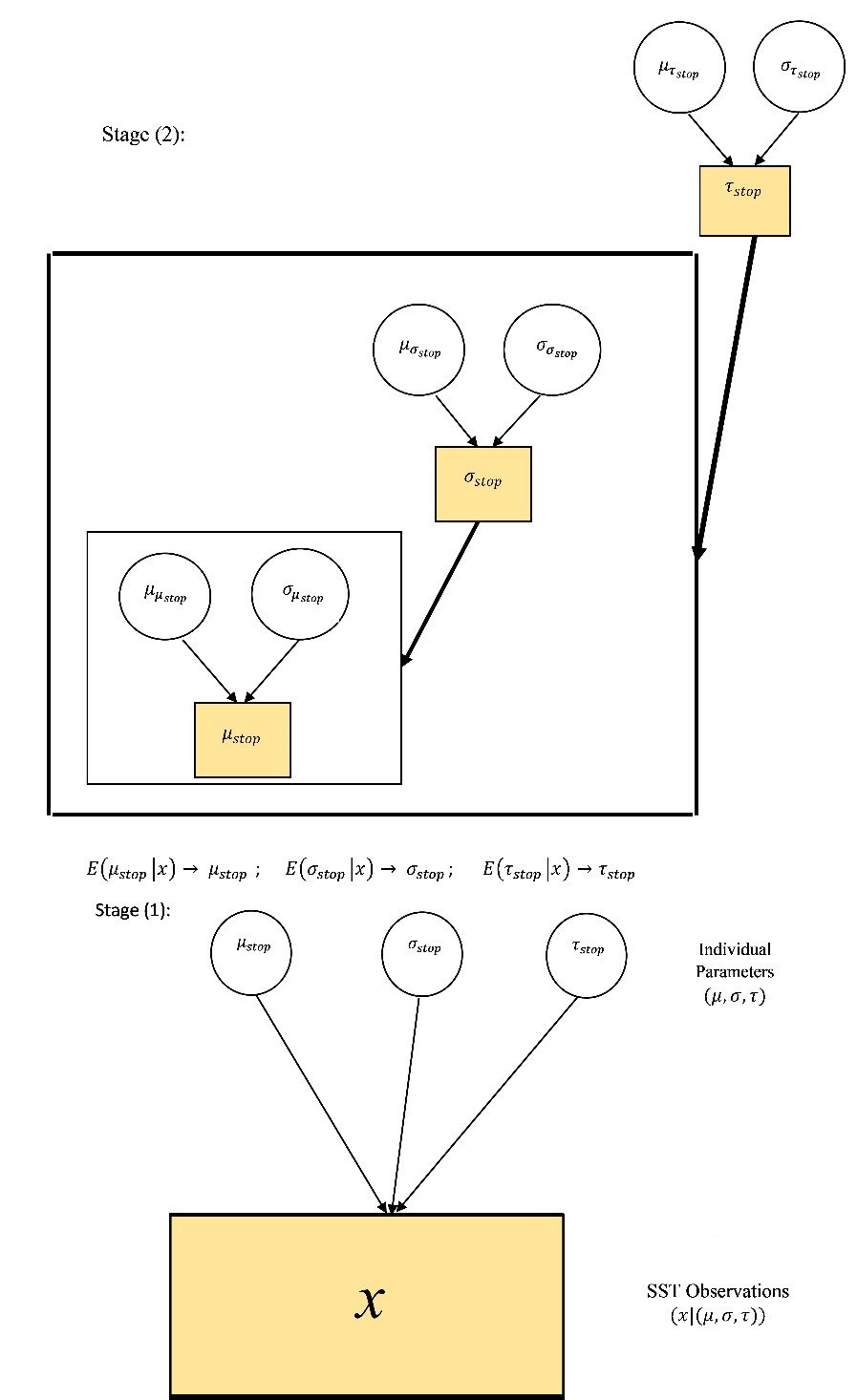}
\caption{Two-Stage Bayesian Parametric Approach (TSBPA) with ExGaussian distributional assumption framework. Filled in shapes indicate known values. }
\end{center}
\end{figure}
\newpage
\textbf{Step(2): Paired Samples Parametric Distribution Test(PSPDT\footnote{This proposed test can be considered as a distributional counterpart of the paired z-test in the real numbers. })}

Using overall estimates in Step (1), we then conduct the following paired samples parametric distribution test hypothesis testing for $K=44$ at the 5\% significance level:\par  
\begin{eqnarray}
H_{0}&:&SSRT_{O.Single}(\overrightarrow{\theta_S})=_{st}SSRT_{O.Mixture}(\overrightarrow{\theta_M}),\nonumber\\
H_{1}&:&SSRT_{O.Single}(\overrightarrow{\theta_S})\neq_{st}SSRT_{O.Mixture}(\overrightarrow{\theta_M}). 
\end{eqnarray}
where the Two-Sample Kolmogorov-Smirnov Statistics $D_{(n,m_k)}$ for the kth $(1\leq k\leq K)$ comparisons of the simulated  distributions in (17),  the following average two-samples KS statistics were considered as the test statistics for the comparison of distributions in the test of (17):\par 
\begin{equation}
\overline{D_{n,m}}=\frac{1}{K}\sum_{k=1}^{K}D_{n,m_{k}}
\end{equation}
We reject the null hypothesis $H_0$ in favor of alternative hypothesis $H_1$ at given $\alpha$-level (e.g., 0.05) whenever:\par 
\begin{equation}
\overline{D_{n,m}}>c(\alpha)\sqrt{\frac{1}{m}+\frac{1}{n}}: c(\alpha)=\sqrt{\frac{-1}{2}ln(\frac{1}{2})}, \alpha=0.05, n=m=96.
\end{equation}
The Two-Sample K-S test analysis was conducted with R3.4.3 software as before.  The hypothesis testing in (17) were repeated for other comparisons between cluster type SSRT indices including  $SSRT_A$  vs $SSRT_{Single}$;  $SSRT_B$  vs $SSRT_{Single}$;and $SSRT_B$  vs  $SSRT_A$.
\begin{remark}
The test (17) for the degenerate case of $K=1$ reduces to the usual two samples K-S test at the individual level (15).
\end{remark} 
\section{Results}
The results are divided into three subsections. In subsection 3.1, we explored the posterior mean Ex-Gaussian parameter estimations of cluster type-A, cluster type-B, single, and mixture SSRT distribution. We compared them for the shape statistics, including skewness and kurtosis across cluster type indices. Next, in subsection 3.2, we compared single SSRT and mixture SSRT distributions in stochastic order at two levels: (i) the individual level; and (ii) the population level. For the individual level, we applied IBPA, and for the population level, we used TSBPA. Finally, in subsection 3.3, we compare the comparison results for the descriptive statistics and the entire SSRT distribution in terms of the cluster weights  ($W_A$). 
\subsection{Descriptive Analysis of Ex-Gaussian Parameters and Shape Statistics}
This section includes two sets of descriptive results: First, the results for cluster-type related parameters, the mean and standard deviation of the Ex-Gaussian SSRT; Second, the results for cluster-type related shape statistics skewness and kurtosis of the involved random variables. Throughout these results, as it is shown in Figure 3, the parameters $(\mu,\sigma,\tau)$ refer to the mean posterior estimates of the random variables $(\mu,\sigma,\tau)$ in the TSBPA, respectively. \par 
Table 2. presents the descriptive results for the type-A, type-B, single, and mixture fitted SSRT Ex-Gaussian random variable using IBPA\footnote{See Supplementary Material (Table 1) for parameter estimates across the three cluster types.}. As it is shown, there is no significant difference between parameters $\theta=(\mu,\sigma,\tau)$, mean and standard deviation between cluster type SSRTs. However, the mentioned list of both cluster types of SSRTs is significantly larger than the single SSRT. Hence, we conclude at this stage that the mean of mixture SSRT is significantly larger than the one of single SSRT. This result is consistent with the frequentist approach \cite{I14}. However, it is observed that the variance has significantly increased, and consequently, the precision has significantly decreased. We remind the reader that there are two evidences for violation of the assumption of equal impact of the preceding trial type (go/stop) on the current stop trial SSRT: First, despite the non-significant results presented in Table 2 (Panel(b): Type B vs. Type A) the mean type-B SSRT has a non-identity linear relationship with mean type-A SSRT ($mean.SSRT_{Bi}=\beta_0+\beta_1.mean.SSRT_{Ai}+\epsilon_i:\epsilon_i\sim N(0,\sigma_{e}^{2}), \beta_0=96.2,$ $ (95\%CI=(4.0,188.4)); \beta_1=0.53(95\%CI=(0.06,1.0))$,). Otherwise, such a relationship must be identity linear; Second,   the mean and standard deviation of type-A SSRT and type-B SSRT are significantly different from those of single SSRT. Otherwise, all these descriptive statistics would have been equal across type-A, type-B, and type-S single SST clusters.\par 

\begin{table}[tbh]
\centering 
\caption{Descriptive results for parameters, mean and standard deviation of fitted IBPA Ex-Gaussian distribution to SSRT given cluster type(n=44).} 
\resizebox{\textwidth}{!}{
\begin{threeparttable}[b]
\begin{tabular}{l cccccc} 
 \hline
 (a) Descriptive Results& & & & & \\  \hline
                  & & Parameter(mean(95$\%$CI))& & Statistics(mean(95$\%$CI))& \\ \hline
Cluster Type & $\mu $ & $\sigma $ & $\tau $ & Mean  & St.d  \\ \hline
Type S & 93.1&116.2&103.6&196.8&
157.8\\
& (83.6,102.7)&(104.1,128.2)&(87.7,119.6)&(173.5,220.1)&
(139.4,176.2)\\
Type A & 124.1&164.5&140.9&265.0&217.7\\
& (107.4,140.7)&(150.4,178.3)&(126.9,155.0)&(235.8,294.2)&(199.1,236.2)\\
Type B & 123.9&168.0&129.7&253.6&213.2 \\
 & 107.2,140.6)&(155.8,180.3)&(115.3,143.9)&(222.9,284.2)&(195.7,231.0) \\  \hline          
(b) Two Sample ttest & & & & & \\  \hline
                  & & Parameter(mean(95$\%$ CI)) &  & Statistics(mean(95$\%$ CI))& \\ \hline
Comparison & $\mu$ & $\sigma$ & $\tau$ & Mean & St.d \\   \hline
Type B vs. Type A & -0.2 & 3.6 & -11.3 & -11.4&-4.4\\
 & (-22.1,21.7) & (-18.1,25.3) & (32.6,10.1) & (-53.5,30.7)&(-33.5,24.7)\\
Type B vs. Type S & 30.8\textsuperscript{***}& 51.7\textsuperscript{***}& 26.0\textsuperscript{***}& 56.8\textsuperscript{***}&55.5\textsuperscript{***}\\
 & (17.9,43.6)& (37.5,66.0)& (10.9,41.1)& (32.2,81.4)&(35.4, 75.7)\\
Type A vs. Type S & 30.9\textsuperscript{***}&48.2\textsuperscript{***}& 37.3\textsuperscript{***}& 68.2\textsuperscript{***}& 59.9\textsuperscript{***}\\
 & (18.5,43.3)&(36.5,59.8)& (24.8,49.8)& (48.3,88.1)& (44.4,75.4)\\
Type M vs. Type S & - & - & - & 63.7\textsuperscript{***}&71.4\textsuperscript{***} \\
 & - & - & - & (57.2,70.1)&(62.5,80.3) \\    \hline
\end{tabular}
\begin{tablenotes}
     \item Notes: $\overline{W_{A}}=0.59,$ ${}^{*}$p-value$<0.05$;${}^{**}$p-value$<0.005$;${}^{***}$p-value$<0.0005$.
 \end{tablenotes}
\end{threeparttable}
}
\end{table} 

Figure 4 shows the difference between skewness and kurtosis of fitted IBPA Ex-Gaussian SSRT random variables by cluster type. As shown in Figure 4 (a)  while each of Mixture SSRT components has smaller or equal skewness versus the Single SSRT, upon combination into Mixture SSRT, the resultant Mixture SSRT has significantly larger skewness compared to the Single SSRT. Similar results hold for the case of kurtosis as shown in Figure 4 (b).\par

\begin{figure}[H]
\begin{center}
\includegraphics[totalheight=14 cm, width=14 cm]{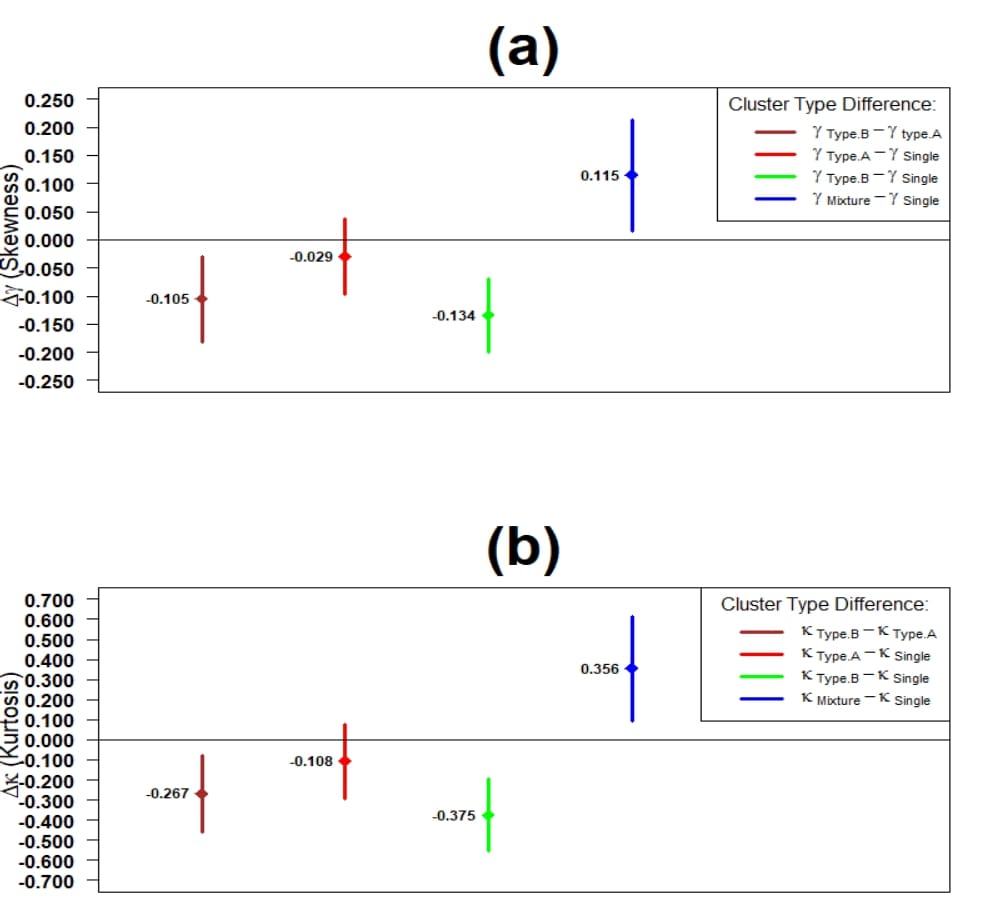}
\caption{ Plot of statistics difference of fitted IBPA Ex-Gaussian SSRT random variable by cluster type (n=44): (a) Skewness;  (b) Kurtosis. }
\end{center}
\end{figure}

Given summary statistics and shape statistics comparison results between single SSRT and mixture SSRT, one naturally considers comparing their associated distributions. In the next section, we deal with this topic.\par  

\subsection{Bayesian Mixture SSRT Estimation and Comparisons}

This section deals with individual and overall level estimations of Single SSRT and Mixture SSRT and their usual stochastic order comparisons. 

\begin{table}[H]
\centering 
\label{table3}
\caption{Two-Sample Kolmogorov–Smirnov test results for the Single SSRT distribution versus Mixture posterior SSRT distribution, (n = 44). }  
\resizebox{\textwidth}{0.75\paperheight}{
\begin{threeparttable}[b]
\begin{tabular}{l cccccc}  
\hline  
 & &Alternative &  &Hypothesis &  &   \\
\hline 
 &Unequal& &Greater & & Less& \\
\hline 
$\#$&  Statistics & p-value& Statistics & p-value& Statistics & p-value \\
\hline 
1& 	0.2708&	0.0017&	0.0417&	0.8465&	0.2708&	0.0009\\
2&	0.2604&	0.0029&	0.0208&	0.9592&	0.2604&	0.0015\\
3&	0.3333&	0.0001&	0.0729&	0.6002&	0.3333&	0.0001\\
4&	0.3438&	0.0001&	0.0417&	0.8465&	0.3438&	0.0001\\
5&	0.2396&	0.0079&	0.0312&	0.9105&	0.2396&	0.0040\\
6& 0.2812&	0.0009&	0.0417&	0.8465&	0.2812&	0.0005\\
7& 0.3021&	0.0003&	0.0729&	0.6002&	0.3021&	0.0002\\
8&	0.2188&	0.0200&	0.0312&	0.9105&	0.2188&	0.0101\\
9&	0.3646&	0.0001&	0.0521&	0.7707&	0.3646&	0.0001\\
10&0.2396&	0.0079&	0.0312&	0.9105&	0.2396&	0.0040\\
11&0.3229&	0.0001&	0.0417&	0.8465&	0.3229&	0.0001\\
12&0.3229&	0.0001&	0.0000&	1.0000&	0.3229&	0.0001\\
13&0.3021&	0.0003&	0.0208&	0.9592&	0.3021&	0.0002\\
14&0.2396&	0.0079&	0.0104&	0.9896&	0.2396&	0.0040\\
15&0.3229&	0.0001&	0.0625&	0.6873&	0.3229&	0.0001\\
16&0.2500&	0.0048&	0.0312&	0.9105&	0.2500&	0.0025\\
17&0.3229&	0.0001&	0.0417&	0.8465&	0.3229&	0.0001\\
18&0.3542&	0.0001&	0.0000&	1.0000&	0.3542&	0.0001\\
19&0.2604&	0.0029&	0.0417&	0.8465&	0.2604&	0.0015\\
20&0.2604&	0.0029&	0.0312&	0.9105&	0.2604&	0.0015\\
21&0.3854&	0.0001&	0.1042&	0.3529&	0.3854&	0.0001\\
22&0.3438&	0.0001&	0.0417&	0.8465&	0.3438&	0.0001\\
23&0.3646&	0.0001&	0.1146&	0.2835&	0.3646&	0.0001\\
24&0.3021&	0.0003&	0.0104&	0.9896&	0.3021&	0.0002\\
25&0.4896&	0.0001&	0.0833&	0.5134&	0.4896&	0.0001\\
26&0.2604&	0.0029&	0.0104&	0.9896&	0.2604&	0.0015\\
27&0.3958&	0.0001&	0.0729&	0.6002&	0.3958&	0.0001\\
28&0.2708&	0.0017&	0.0208&	0.9592&	0.2708&	0.0009\\
29&0.2396&	0.0079&	0.0104&	0.9896&	0.2396&	0.0040\\
30&0.3750&	0.0001&	0.0729&	0.6002&	0.3750&	0.0001\\
31&0.4062&	0.0001&	0.0312&	0.9105&	0.4062&	0.0001\\
32&0.2500&	0.0048&	0.0208&	0.9592&	0.2500&	0.0025\\
33&0.2500&	0.0048&	0.0208&	0.9592&	0.2500&	0.0025\\
34&0.1562&	0.1923&	0.0000&	1.0000&	0.1562&	0.0960\textsuperscript{*}\\
35&0.2500&	0.0048&	0.0104&	0.9896&	0.2500&	0.0025\\
36&0.3021&	0.0003&	0.0312&	0.9105&	0.3021&	0.0002\\
37&0.1979&	0.0463&	0.0104&	0.9896&	0.1979&	0.0233\\
38&0.3125&	0.0002&	0.0312&	0.9105&	0.3125&	0.0001\\
39&0.2917&	0.0005&	0.0625&	0.6873&	0.2917&	0.0003\\
40&0.3750&	0.0001&	0.0521&	0.7707&	0.3750&	0.0001\\
41&0.2188&	0.0200&	0.0417&	0.8465&	0.2188&	0.0101\\
42&0.3021&	0.0003&	0.0208&	0.9592&	0.3021&	0.0002\\
43&0.2188&	0.0200&	0.0104&	0.9896&	0.2188&	0.0101\\
44&0.4062&	0.0001&	0.0312&	0.9105&	0.4062&	0.0001\\
\hline
\end{tabular}
\begin{tablenotes}
     \item \small{Notes: IBPA: $\#$Chains = 3; Simulations = 20,000; Burn-in = 5,000 (for both single and mixture parameters); The sample size for K-S test for each distribution was n=m=96; $^*:$ Exceptional case.}  
 \end{tablenotes}
\end{threeparttable}
}
\end{table} 

Table 3 presents the results of the two-sample KS hypothesis test at the individual level given by (15) by direction and p-values for the sample of 44 subjects based on IBPA. Similar hypothesis testing is conducted replacing $\neq$ with $<$  and $>$ in an alternative test. With one participant's exception (case 34),  the result shows that the single SSRT is smaller than the mixture SSRT in stochastic order. This result is consistent with the direction of constant index SSRT results \cite{I14}.\par 
We test the hypothesis (17) from  TSBPA given uninformative priors for an overall conclusion using a paired samples parametric distribution test. As before, similar hypothesis testing is conducted by replacing $\neq$ with $<$  and $>$ in the alternative test. The choice of TSBPA rather than HBPA was out of consideration for pairwise non-zero correlations in the second stage of the analysis. One key missing characteristic in the HBPA is the relaxing assumption of zero correlation of mean posterior parameters at the individual level. This assumption  is violated given cluster-S SSRT mean posterior parameters Pearson correlations   $\rho_{\mu\sigma}=0.20, \rho_{\mu\tau}=0.64,\rho_{\sigma\tau}=0.66;$ cluster-A SSRT mean posterior parameters Pearson correlations $\rho_{\mu\sigma}=0.52, \rho_{\mu\tau}=0.81,\rho_{\sigma\tau}=0.74;$  and, cluster-B SSRT mean posterior parameters Pearson correlations $\rho_{\mu\sigma}=0.69, \rho_{\mu\tau}=0.95,\rho_{\sigma\tau}=0.80$.  Table 4 presents the results of the paired samples parametric distribution test using TSBPA: 

\begin{table}[!htbp]
\centering 
\label{table4}
\caption{Two-Sample Kolmogorov–Smirnov test results for the Single SSRT distribution versus the Mixture SSRT distribution, (n = 44).}  
\resizebox{\textwidth}{!}{
\begin{threeparttable}[b]
\begin{tabular}{l  cccccc}  
\hline  
 & &Alternative &  &Hypothesis &  &   \\
\hline 
  &Unequal& &Greater & & Less& \\
\hline 
Comparison & Statistics & p-value& Statistics & p-value& Statistics & p-value \\
\hline 
$SSRT_{Single}$ vs. $SSRT_{Mixture}$ &0.2100& $<$0.0302 & 0.0736 & $>$0.9999 & 0.2095 & $<$ 0.0312 \\
$SSRT_{Single}$ vs. $SSRT_{B}$ & 0.2017 & $<$0.0422 & 0.0816 & $>$0.9999 & 0.1984 & $<$0.0562\\
$SSRT_{Single}$ vs. $SSRT_{A}$ & 0.2273 & $<$0.0152 & 0.0658 & $>$ 0.9999 & 0.2256 & $<$0.0152\\
$SSRT_{B}$ vs. $SSRT_{A}$ & 0.1326 & 0.3702 & 0.1075 & 0.6502 & 0.0653 & $>$ 0.9999 \\ \hline
\end{tabular}
\begin{tablenotes}
     \item Note: The sample sizes for K-S test for both distributions were $n=m=96$. $\overrightarrow{\theta_S}=(\overline{\mu_T},\overline{\sigma_T},\overline{\tau_T})=(78.4,93.9,73.1),$ $\overrightarrow{\theta_A}=(\overline{\mu_A},\overline{\sigma_A},\overline{\tau_A})=(94.0,134.5,104.8),$ $\overrightarrow{\theta_B}=(\overline{\mu_B},\overline{\sigma_B},\overline{\tau_B})=(90.9,142.3,99.0)$ and $\overrightarrow{\theta_M}=(\overline{W_A},\overrightarrow{\theta_A},\overrightarrow{\theta_B}),$ with $\overline{W_{A}}=0.59.$ 
 \end{tablenotes}
\end{threeparttable}
}
\end{table}

As we observe from Table 4, the results are conclusive. The single SSRT is (provisionally) smaller than cluster type-A SSRT, cluster type-B SSRT (p-value $<$0.0562); and, Mixture SSRT. Also, given consideration of mean of SSRT distribution as its point index estimation, the result regarding the comparison of single SSRT versus Mixture SSRT is consistent with the direction of the frequentist results as:  
$$SSRT_{Single}^{c}=E(SSRT_{Single})=151.5<195.2=E(SSRT_{Mixture})=SSRT_{Weighted}^c.$$
Note that there is no significant difference between cluster type A and cluster type B SSRT (p-value=0.370). We remind the reader that one may guess that cluster type-A SSRT is smaller than the cluster type-B SSRT in stochastic order given the test p-value 0.650 versus  0.999, but this claim's verification needs a much larger sample size than n=44.\par 
Figure 5. shows the plot of the overall density and cumulative distribution function of cluster type SSRTs with overall TSBPA parameter estimates given in Table 4. As it is observed in Figure 5(b), while there is no such distinction between cumulative distributions of cluster type-A SSRT, cluster type-B SSRT, and Single SSRT, the cumulative distribution of single SSRT is clearly on the left side of that of Mixture SSRT.\par 
\begin{figure}[H]
\begin{center}
\includegraphics[totalheight=14 cm, width=14 cm]{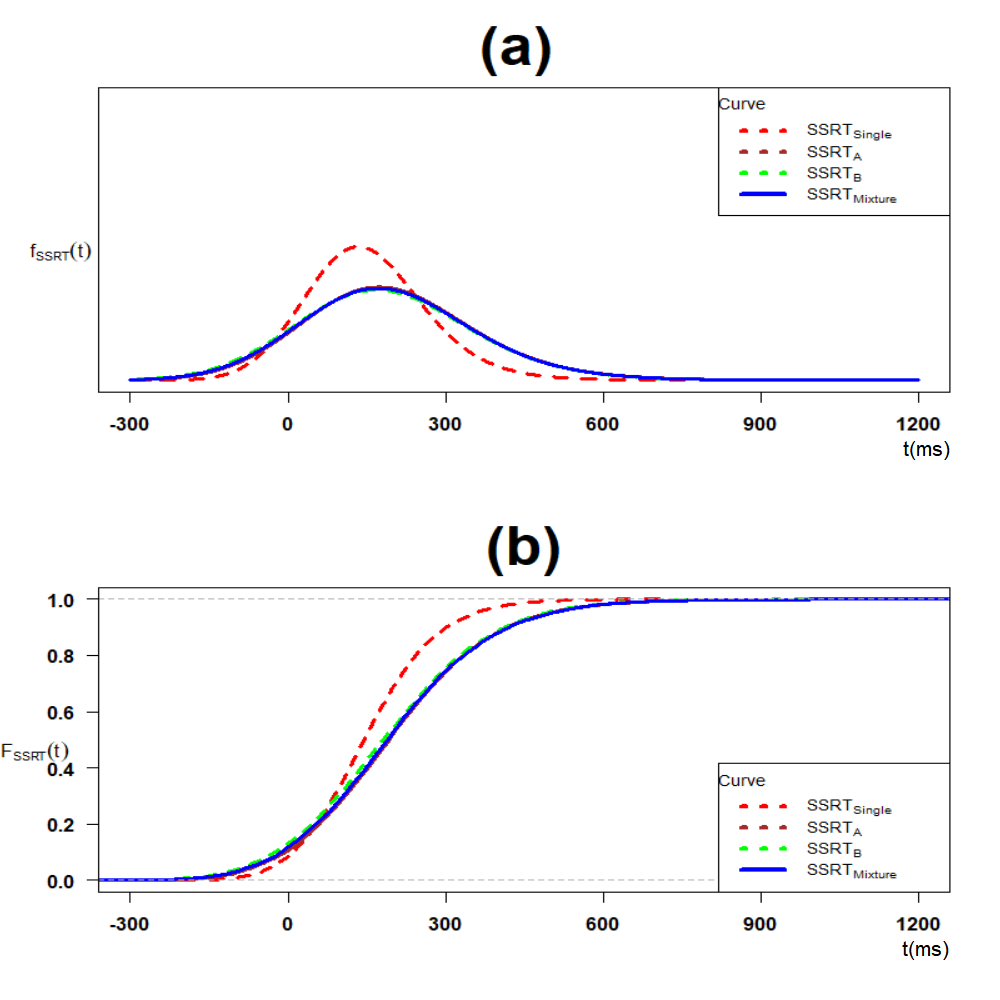}
\caption{ The density and cumulative distribution function(CDF) of overall sample cluster type SSRT, single SSRT and mixture SSRT with Ex-Gaussian parametric distribution: (a) density; (b) cumulative distribution function.}
\end{center}
\end{figure}
In this and the previous section, we considered the cluster type weight $(W_A)$ in its fixed individual values. In the next section, we study its role in the comparison results as a critical variable on its own.\par 
\newpage
\subsection{The Role of Cluster Type Weights in the Comparisons}

This section compares the descriptive statistics of mean, variance, and the entire distribution of SSRT indices in terms of individual optimal weights.  By definition, the optimal weight $W_A$ is the most natural weight given the independence of assignment of stop or go process to the given trial  \cite{I15}.    The following proposition determines the values of the optimal weight \cite{I15}:
\begin{proposition}
The weight $W_A = 0.75(W_B = 0.25)$ is the optimal weight given independence of assignment of stop or go process to the given trial in the tracking SST data with proportion of 25\% of stop trials.
\end{proposition}
Note that the fitted ExG parameters $\theta=(\mu,\sigma,\tau)$ in each cluster type SST data are independent of the weights $W_A.$ This result is because the fitted ExG parameters for the  SSRT are independent of the stop trials' proportion. Hence, from the weight $W_A$ (as the result of the equality in Proposition 3.1.). Given this result, we discuss the impact of cluster type weights on average disparities of mean SSRT estimates and the variance SSRT estimates as follows: \par  
First, to study the impact of individual weights on the disparities of the mean estimations across indices, we consider average differences of the new index $SSRT_{Mixture}$  mean versus the established index $SSRT_{Single}$   mean in terms of the individual weights $(W_A)$. The averages of the ExG parameters are taken over entire $n=44$ participants. Considering $W_A$ as the main variable, it follows that: 

\begin{eqnarray}
\overline{\Delta E(SSRT)}(W_A)&=& \overline{E(SSRT_{Mixture})}-\overline{E(SSRT_{Single})}\nonumber\\
&=& \overline{E(W_A.SSRT_{A}+(1-W_A).SSRT_{B})}-\overline{E(SSRT_{Single})}\nonumber\\
&=& W_A.\overline{E(SSRT_A)}+(1-W_A).\overline{E(SSRT_B)}-\overline{E(SSRT_{Single})}\nonumber\\
&=& (\overline{E(SSRT_A)}-\overline{E(SSRT_B)}).W_A+ (\overline{E(SSRT_B)}-\overline{E(SSRT_{Single})} \nonumber\\
&=& ((\overline{\mu_A}+\overline{\tau_A})-(\overline{\mu_B}+\overline{\tau_B})).W_A + ((\overline{\mu_B}+\overline{\tau_B})-(\overline{\mu_S}+\overline{\tau_S})) \nonumber\\
&=&11.4\ \  W_{A}+ 56.8,\ \ \hspace{3 cm} (0\leq W_A \leq 1).
\end{eqnarray}

Figure 6(a) presents the average disparities of mean mixture SSRT and mean single SSRT   versus individual weights $W_A$ for the extrapolated range of $[0,1].$ As shown,  the average difference between two index means is linear in terms of the individual weight $W_A$ with the minimum value of 56.8 ms (for minimum sample weight of 0.00) and maximum value of 68.2 ms (for maximum sample weight of 1.00). Also, their corresponding averaged disparities equals 63.5 ms (at the overall sample weight of 0.59 ). Finally, the two index means' disparities are maximized to 65.3 ms using the optimal weight of 0.75.\par
   
Second, to examine the impact of individual weights on the disparities of the variance estimations across indices, similar to the case in section 3.2, we consider average differences of the new index $SSRT_{Mixture}$  variance versus the established index $SSRT_{Single}$   variance in terms of the individual weights $(W_A).$ The averages of the quadratic ExG parameters are taken over entire $n=44$ participants. Considering $W_A$ as the primary variable, it follows that: 
\newpage 
\begin{eqnarray}
\overline{\Delta Var(SSRT)}(W_A)&=& \overline{Var(SSRT_{Mixture})}-\overline{Var(SSRT_{Single})}\nonumber\\
&=& \overline{Var(W_A.SSRT_{A}+(1-W_A).SSRT_{B})}-\overline{Var(SSRT_{Single})}\nonumber\\
&=& -\overline{(E(SSRT_A)-E(SSRT_B))^2}. W_{A}^{2}\nonumber\\
&& + \bigg(\overline{(E(SSRT_A)-E(SSRT_B))^2}\nonumber\\
&&+\overline{(Var(SSRT_A)-Var(SSRT_B))}\bigg).W_{A}\nonumber\\
&& + \overline{Var(SSRT_{B})-Var(SSRT_{Single})}\nonumber\\
&=& -\overline{(\mu_A+\tau_A-\mu_B-\tau_B)^2}.W_{A}^{2}\nonumber\\
&& + (\overline{(\mu_A+\tau_A-\mu_B-\tau_B)^2}+\overline{(\sigma_{A}^{2}+\tau_{A}^{2}-\sigma_{B}^{2}-\tau_{B}^{2})}).W_{A} \nonumber\\
&& \overline{\sigma_{B}^{2}+\tau_{B}^{2}-\sigma_{S}^{2}-\tau_{S}^{2}} \nonumber\\
&=& -18885.9\ \ W_{A}^{2} + 21106.8\ \  W_{A} + 20330.3,\hspace{1 cm}  (0\leq W_A \leq 1). \nonumber\\
&&
\end{eqnarray}
In particular, the average variance differences attains its maximum value at:
$$W_{A}=\frac{\overline{(\mu_A+\tau_A-\mu_B-\tau_B)^2}+\overline{(\sigma_{A}^{2}+\tau_{A}^{2}-\sigma_{B}^{2}-\tau_{B}^{2})}}{2\overline{(\mu_A+\tau_A-\mu_B-\tau_B)^2}}\approx 0.56.$$

Figure 6 (b) presents the average disparities of mixture SSRT variance and single SSRT variance versus individual weights $W_A$ for the extrapolated range of $[0,1].$ As shown,  the average difference between two indices variances follows a quadratic increasing- decreasing pattern in terms of the weights $W_{A}$ with maximum values attained closed to $W_A\approx 0.56.$ Next, the disparities for the optimal weight $W_A=0.75$ is smaller than that of population weight $W_A=0.59.$  However, across all weights spectrum, the average SSRT variance differences are positive, showing that the new mixture SSRT index has higher variance than the current single SSRT index. Consequently, its precision is smaller. \par

Finally, to explore the impact of cluster type weights $(W_A)$ on the overall SSRT distributions comparison results for the hypothesis testing (17), we considered the averaged two-sample KS test statistics S as a function S of weights $S=S(W_A)$ and calculated the corresponding p-values. Figure 7 presents the results in terms of the weights. As shown in figure 7(b), for almost all ranges of the weights $W_A$, the single SSRT is significantly smaller than the mixture SSRT in stochastic order. Next, the disparity is the weakest when $W_A$=0 with the corresponding p-value=0.0562, and it is the strongest when $W_A$=1 with the corresponding p-value=0.0152. Finally, the disparity at the optimal weight $W_A$=0.75 is more potent than that of population weight $W_A$=0.59 with corresponding p-values of 0.0262 and 0.0312, respectively. \par 

\begin{figure}[H]
\begin{center}
\includegraphics[totalheight=14 cm, width=14 cm]{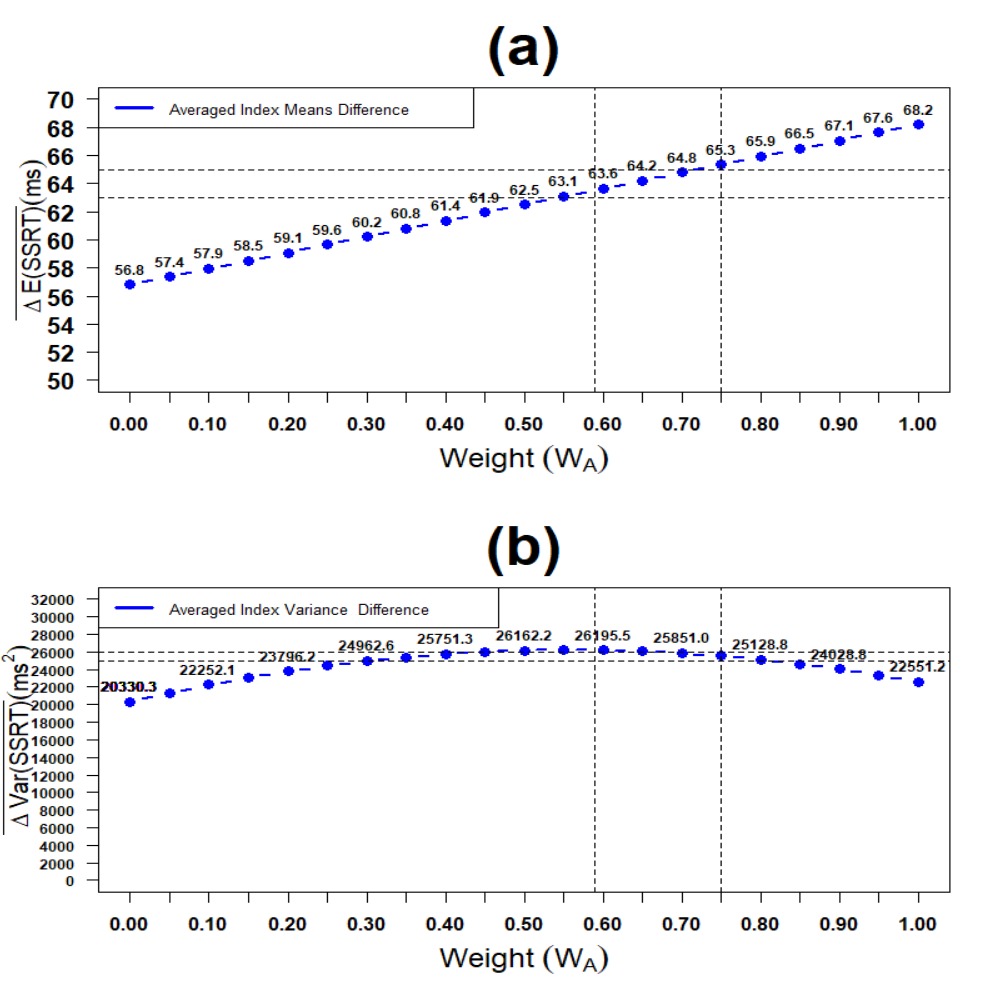}
\caption{Plot of difference of SSRT index statistics ($SSRT_{Mixture}$ vs $SSRT_{Single}$)  by weight $W_A(n=44)$: (a) Means; (b) Variances}
\end{center}
\end{figure}
\begin{figure}[H]
\begin{center}
\includegraphics[totalheight=14 cm, width=14 cm]{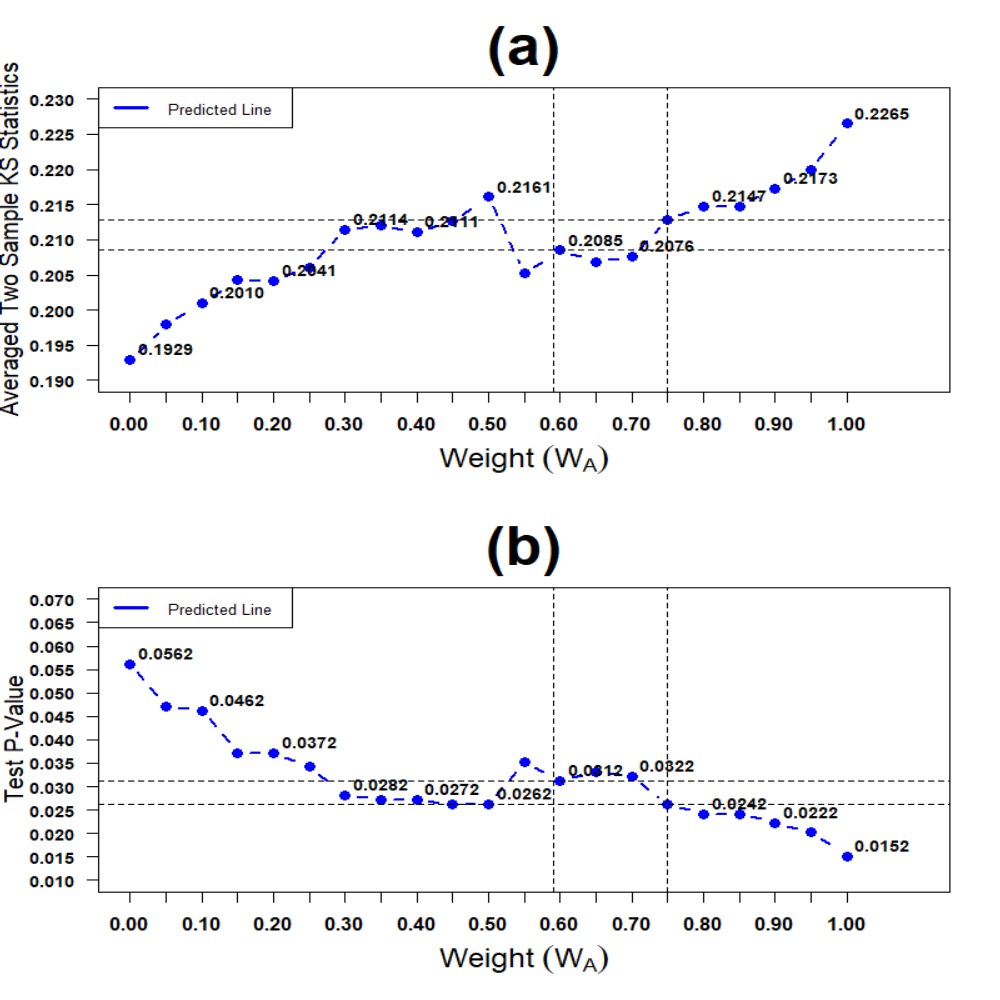}
\caption{Plot of overall test results of single SSRT versus mixture SSRT : (a) Two sample KS test statistics; (b) cut off point of the test p-value }
\end{center}
\end{figure}

\section{Discussion}

\subsection{Present Work}

This study presented a mixture Bayesian parametric approach for a more illuminating SSRT distribution estimation by considering two subtype SST cluster information suggesting a new estimation of the SSRT distribution. Also, it introduced two novel statistical methodologies accompanied by their empirical applications: TSBPA and PSPDT. It was hypothesized that considering cluster type information in the new mixture SSRT distribution calculations would impact the estimation of SSRT distribution. This yields to a distributional counterpart to the case of constant index SSRT \cite{I14, I15}.\par  
The results confirmed the hypothesis through three observations:\newline\\
\begin{enumerate}
\item The descriptive and shape statistics
\item The distributional comparisons at the individual level and the population level
\item With the validity of the results in the first two observations across the entire spectrum of the weights
\end{enumerate}
Similar to the constant index SSRT \cite{I14}, in most cases, the mixture SSRT is different from the single SSRT in shape statistics and the stochastic order. However, in two special distinct cases, they are the same: (i) type A cluster SST is empty $(W_A=0)$ and (ii) type B cluster SST is empty $(W_A=1).$\par  
This study confirmed that SSRT depends on non-horse race-related factors in each round of SST experimental trial, such as memory aftereffects and proportion of cluster type stop trials. It has shed light on the preparation aspect of choice stop-signal reaction times by treating the previous trial type's aftereffects as memory in the two-state mixture model \cite{I30}. Besides, given that skewness of the RT distributions increases with memory involvement (versus a perceptual decision)\cite{D1}, the increase in reported skewness in the mixture SSRT versus single SSRT confirms that the proposed mixture model successfully captures the memory involvement in the decision process\cite{D1}. Also, as in the context of the horse race independent model, an increase in kurtosis of SSRT is proportional to more extreme values of the right tail of SSRT distribution.  Hence, this causes a higher probability of failed inhibition in the stop trials(and vice versa). Next, the increase in reported kurtosis in the mixture SSRT versus single SSRT gives evidence that the proposed mixture model optimally uses the information given by pre pushed failed inhibitions in the stop trials in the estimation of SSRT distribution. \par 
This study's findings for the SSRT distribution were consistent with the constant index SSRT when considering the impact of sub-cluster types in the estimations \cite{I14,I15}.  In detail, there were consistent results between the usual comparison of the single SSRT and weighted SSRT (as constant indices) and the stochastic comparison of the single SSRT and mixture SSRT (as non-constant random variables). Indeed, we found that:
\begin{eqnarray}
SSRT_{Single}^{c}&<&SSRT_{Weighted}^{c},\\
SSRT_{Single} &<_{st}&  SSRT_{Mixture}.
\end{eqnarray}
On the one hand, if we look at two sides of the equation(22) as degenerate random variables, we are led to the equation (23). On the other hand, if we take expectations from both sides of the equation (23), we are led to the equation (22). \par 
The study's novel statistical methodological contributions involved the Two-Stage Bayesian Parametric Approach (TSBPA) and the Paired Samples Parametric Distribution Test(PSPDT). TSBPA's advantage was that it considers the underlying non-zero correlation between estimated mean posterior parameters at the first stage in the second stage's final calculations. This feature is neglected in HBPA. PSPDT offered a novel method to compare paired sets of parametric random variables using the two-samples KS test. An application of both proposed methods was provided in this study. \par  
There are limitations in the current study. First, the sample size was relatively small $(n=44).$ To show more precise comparisons, larger sample size is needed. Second, the TSBPA assumes a multivariate normal distribution form for the mean posterior parameters at the second stage, which may not hold. Third, in TSBPA, when comparing overall Mixture SSRT and overall Single SSRT, there is no specific restriction on the simulation sample sizes in equation (18). Here, while we set the sample sizes to $n = m = 96$ (the SST data trials size), there could be other choices.  Finally, given the structure of the equations for the shape statistics (skewness and kurtosis) in terms of the cluster weights $(W_A),$ unlike the descriptive statistics in section 3.3, there was no simple closed form for the averaged differences of the new index $SSRT_{Mixture}$  skewness (or kurtosis) versus the established index $SSRT_{Single}$ skewness (or kurtosis) in terms of the individual weights $(W_A).$ Similar to the descriptive statistics, the existence of such a simple closed formula would shed more light on the average disparities of skewness and kurtosis of the two indices across a spectrum of the individual weights. \par

\subsection{Future Work}

The proposed approach in modelling the SSRT distribution in this study should be replicated in future research in several different directions. This further work may shed light on further unknown corners. New work includes: (i)considering the larger number of SST trials,(ii) examining the order of trials,(iii)expanding these methods to other clinical populations,(iv) considering trigger failures in the modelling,(v) interpreting the shape statistics, and (vi) estimating signal respond reaction times (SRRT).\par  
First, research has recommended that reliable estimates of SSRT for adults requires 200 SST trials with 50 stops \cite{I16}. Hence,  the current work's approach needs to be replicated for SST data with 400 trials, including sub-cluster types of 200 trials with 50  stops for confirmation and generalization purposes.\par 
Second, additional research to this study must address the presumption of equal impact in the order of trials for the same cluster type weights $W_A.$ For example, for the case of  $W_A=1,$  one may consider two schemes within the study of 96 SST trials: In the first scheme, trials numbered $2k   (1\leq k\leq 25)$ are stop trials. In the second scheme, trials numbered    $98-2k\ (1\leq k\leq 25)$ are the stop trials. There is no known study investigating if, in the same participant, these schemes lead to the same $SSRT_{Mixture}$ or not.\par  
Third, after this study, the work should apply the proposed $SSRT_{Mixture}$ to study the inhibitory deficiency in different clinical groups such as ADHD, OCD, autism, and schizophrenia. The application is in terms of descriptive statistics, shape statistics, plus the differential disparities across these clinical groups.\par   
Fourth, there are trigger failures that impact the estimations \cite{I26-2}. Given the probability of trigger failures(TF) of $P_T(TF);$ $P_A(TF)$ and $P_B(TF)$ for the overall SST data, cluster-A SST data, and cluster-B SST data, respectively, there remains an open question on their relationships and the impact of the cluster type trigger failures in the estimations of the $SSRT_{Mixture}$ and on the above results. The results of such consideration generalize this study's findings in terms of trigger failures and control them in order to eliminate a potential confounding variable, trigger failure status.\par 
Fifth,  this study merely reported and compared the shape statistics for skewness and kurtosis across the cluster type SSRTs, single SSRT, and the mixture SSRT distributions. There is a need to investigate these shape statistics' psychiatric and psychopathological interpretations given the  Ex-Gaussian parametric distribution assumption. \par 
Finally, this and the earlier study \cite{I14} addressed the estimation of stop signal reaction times (SSRT) in the case of the violated assumption of similar aftereffects of the prior trial type. It is plausible to conduct a counterpart investigation on to the estimation of the signal respond reaction times (SRRT) constant index and distribution.\par

\subsection{Conclusion}

There has been a great deal of interest in the aftereffects of inhibition on the estimation of SSRT in the SST literature from the early 1990s. This study addressed the problem in part and presented a two-state mixture model of SSRT distribution by considering the prior trial type with results consistent with the constant SSRT index results in the literature \cite{I14}. The results were consistent across constant index and non-constant random variable contexts in term of the algebraic directions of the comparisons. Moreover, more information was used from the same SST data in the newly proposed mixture estimation method versus the current single estimation method. The vital assumption introduced in this work was relaxed in the newly proposed mixture estimation method. Given these advantages, the researchers are recommended considering mixture SSRT distribution $(SSRT_{Mixture})$  as the most informative estimation of the latency of stopping.

\section*{Acknowledgments}
The authors are grateful to the reviewer for offering constructive comments and posing enriching questions on the first draft of the manuscript.

\section*{Disclosure Statement}
This work has  previously been presented as a research poster in Statistical Society of Canada 2018 annual meeting in Montreal, Canada. Mohsen Soltanifar, Michael Escobar and Annie Dupuis have no financial interests to disclose.  Russell Schachar has equity in ehave and has been the on the Scientific Advisory Board(SAB) in Lilly and Highland Therapautic Inc, Toronto, Canada.
\section*{Funding}
This work has been funded by University of Toronto’s Queen Elizabeth II (QEII) graduate scholarship in science and technology 2017-2018 and, by Canadian  Institute of Health Research (CIHR) operating grant (PIs: Russell Schachar: MOP-93696; Paul Arnold: MOP-106573).
\section*{Note on Contributors}

\textit{Mohsen Soltanifar (Ph.D)} was a sessional course instructor jointly working at the University of Toronto and the Hospital of Sick Children, Toronto. His research area includes quantitative psychology with statistical focus on multilevel, frequentist and bayesian mixture modelling and time series.\newline\\
\textit{Michael Escobar (Ph.D)} is a professor of biostatistics at the University of Toronto. His research interests include nonparametric bayesian modelling, mixture modelling and applied statistical methods in psychiatric research. \newline\\
\textit{Annie Dupuis (Ph.D)} is an adjunct professor of biostatistics at the University of Toronto and an independent biostatistical consultant in Toronto, Canada . Her research areas include attention deficit hyperactivity disorder, autism spectrum disorders, and cystic fibrosis.\newline\\
\textit{Russell Schachar (M.D)} is a  professor of psychiatry at the University of Toronto, and a senior scientist  leading a cognitive neurosciences laboratory in the Research Institute at the Hospital for Sick Children, Toronto. He is the TD Bank chair in Child and Adolescent psychiatry. His research focuses on attention deficit hyperactivity disorder.

\section*{ORCID}

Mohsen Soltanifar:\ \ https://orcid.org/0000-0002-5989-0082\newline
Michael Escobar: \ \ \   https://orcid.org/0000-0001-9055-4709\newline
Annie Dupuis:\ \  \ \ \ \ \         https://orcid.org/0000-0002-8704-078X\newline
Russell Schachar:\ \ \   https://orcid.org/0000-0002-2015-4395\newline

\section*{Abbreviations}
$
\begin{array}{ll}
\text{ADHD} & \text{Attention Deficit Hyperactivity Disorder} \\
\text{BEESTS} & \text{Bayesian Ex-Gaussian Estimation of Stop Signal RT distributions}\\
\text{BPA}& \text{Bayesian Parametric Approach}\\
\text{ExG} & \text{Ex-Gaussian distribution} \\
\text{GORT} & \text{Reaction Time in a go trial}\\
\text{GORTA} & \text{Reaction Time in a type A go trial}\\
\text{GORTB} & \text{Reaction Time in a type B go trial}\\
\text{HBPA} & \text{Hierarchical Bayesian Parametric Approach}\\
\text{IBPA} & \text{Individual Bayesian Parametric Approach}\\
\text{MCMC} & \text{Marco Chain Monte Carlo}\\
\text{OCD} & \text{Obsessive Compulsive Disorder}\\
\text{PSPDT}&\text{Paired Samples Parametric Distribution Test}\\
\text{SSD} & \text{Stop Signal Delay}\\
\text{SRRT} & \text{Reaction Time in a failed stop trial}\\
\text{SRRTA} & \text{Reaction Time in a failed type A stop trial}\\
\text{SRRTB} & \text{Reaction Time in a failed type B stop trial}\\
\text{SSRT} & \text{Stop Signal Reaction Times in a stop trial}\\
\text{SSRTA} & \text{Stop Signal Reaction Times in a type A stop trial}\\
\text{SSRTB} & \text{Stop Signal Reaction Times in a type B stop trial}\\
\text{SST} & \text{Stop Signal Task }\\
\text{TF} & \text{Trigger Failure}\\
\text{TSBPA}& \text{Two Stage Bayesian Parametric Approach}
\end{array}
$

\section*{Supplementary Materials}
Additional supporting information may be found in the online version of this article at the publisher's website on the data of Mean posterior Ex-Gaussian parameters estimations across trial types by IBPA (n = 44).

\clearpage




\end{document}